%% file: Replyv2.tex
\documentclass[aps,prd,amsmath,twocolumn,longbibliography]{revtex4-1}

\usepackage{graphicx}
\usepackage{amssymb,color}
\def\Dslash{\raise.15ex\hbox{/}\kern-.7em D}
\def\Pslash{\raise.15ex\hbox{/}\kern-.7em P}

\newcommand{\ben}{\begin{displaymath}}
\newcommand{\een}{\end{displaymath}}
\newcommand{\be}{\begin{equation}}
\newcommand{\ee}{\end{equation}}
\newcommand{\bea}{\begin{eqnarray}}
\newcommand{\eea}{\end{eqnarray}}

\newcommand{\beqn}{\begin{equation}}
\newcommand{\eeqn}{\end{equation}}

\begin{document}

 NT@UW-23-12 \title{
Comment on ``Electromagnetic form factors for nucleons in short-range correlations": arXiv:2305.13666 }

\author{Dmitry N. Kim$^1$, Or Hen$^2$,  Gerald A. Miller$^1$,  E. Piasetzky$^3$,  M. Strikman$^4$ and L. Weinstein$^5$}

\affiliation{$^1$ 
Department of Physics,
University of Washington, Seattle, WA 98195-1560, USA}

 \affiliation{$^2$
Department of Physics,
Massachusetts Institute of Technology, Cambridge Massachusetts 02139, USA }
        
        \affiliation{$^3$ 
School of Physics and Astronomy, Tel Aviv University, Tel Aviv 69978, Israel}

\affiliation{ $^4$
Department of Physics,
The Pennsylvania State  University,   Norfolk, Virginia, 23529, USA}
\affiliation{ $^5$
Department of Physics,
Old Dominion University,   Norfolk, Virginia, 23529, USA }
                                                                           
\date{\today}     
\maketitle     
%\textbf{Just fixed some minor typos and rewrote some sentences. Also changed the Figure caption to be more complete. Everything that was said in this text makes sense to me. Also there are two acknowledgements sections, one of them should be removed.  }

      Recent experimental studies~\cite{Weinstein:2010rt,Hen:2016kwk,Duer:2018sxh,Schmookler:2019nvf}   seem to support models in which the EMC effect is mainly due to the  influence of nucleon-nucleon short-range correlations (SRC)   (high-relative momentum nucleon-nucleon correlations). %may be a major contributor to the nuclear EMC effect. 
      This hypothesis requires that the structure function for nucleons involved in short-range correlations be heavily suppressed compared to that of free nucleons.
  
      In Ref.~\cite{Xing:2023uhj} the authors use
        calculations performed within an AdS/QCD motivated, light-front quark-diquark model to argue that this large suppression of the nucleon structure function leads to a strong suppression ({\it e.g.} by about 50 \% at $Q^2=1 $ GeV$^2$)   of the nucleon elastic form factors.  This result would seem to imply that  the strong correlation   between SRC and the EMC effect is accidental, because the medium modification of  the form factors has been known to be small~\cite{Sick:1985ygc}, with an upper limit between 3 and 6\% for values of $Q^2$ between 1 and 4 GeV$^2$ for values of missing momentum less than about 150 MeV/c.

  This paper presents a counter-example, showing that a large modification of form factors is not a model-independent result of SRC effects, as   the medium modifications of form factors can be  small.  
  This is done by      showing that a previously published model of the EMC effect~\cite{Kim:2022lng}, also performed using an AdS/QCD motivated light-front quark-diquark model, motivated by the  notion of nucleonic point-like configurations~\cite{Frankfurt:1985cv},  leads to very modest modification of in medium elastic electromagnetic form factors. That model requires bound nucleons to be significantly off the mass-shell  ({\it i.e.} $|p^2-M^2|/M^2$ is far from zero~\cite{CiofidegliAtti:2007ork}), an effect that is a signature for the existence of SRCs, to reproduce nuclear deep inelastic scattering data. The published paper already shows that the slope of the electromagnetic form factors is much less modified than the current experimental uncertainties. The calculations are extended here  to exhibit  the $Q^2$ dependence of the in-medium form factors.  
        
        The free and medium-modified elastic form factors $F_1$, as determined in~\cite{Kim:2022lng}  are shown in Fig.~\ref{Mmod} for a Pb nucleus where the EMC effect is largest, providing an upper limit within the model. Note that the model uses an average virtuality. The influence of the nuclear medium  is shown to be very modest indeed. The published model predicts no medium modification in the form factor $F_2$. Thus the model predicts medium-modifications that are smaller than any existing measurements. %-{\it Mark supply refs for your email of 5/25}
        
        The basic reason for this is that the measured  EMC effect  is no more than 10-15\% and  %The work of Ref.~\cite{Xing:2023uhj} shows no plots to compare with measured nuclear deep inelastic scattering data, but the model of Ref.~\cite{Kim:2022lng}  is  explicitly constructed to reproduce the data.  
       within the model the probability that a given bound nucleon has an excited state component is proportional to the square of the size of the EMC effect.

\begin{figure}[h]
  \centering
 \includegraphics[width=0.5\textwidth]{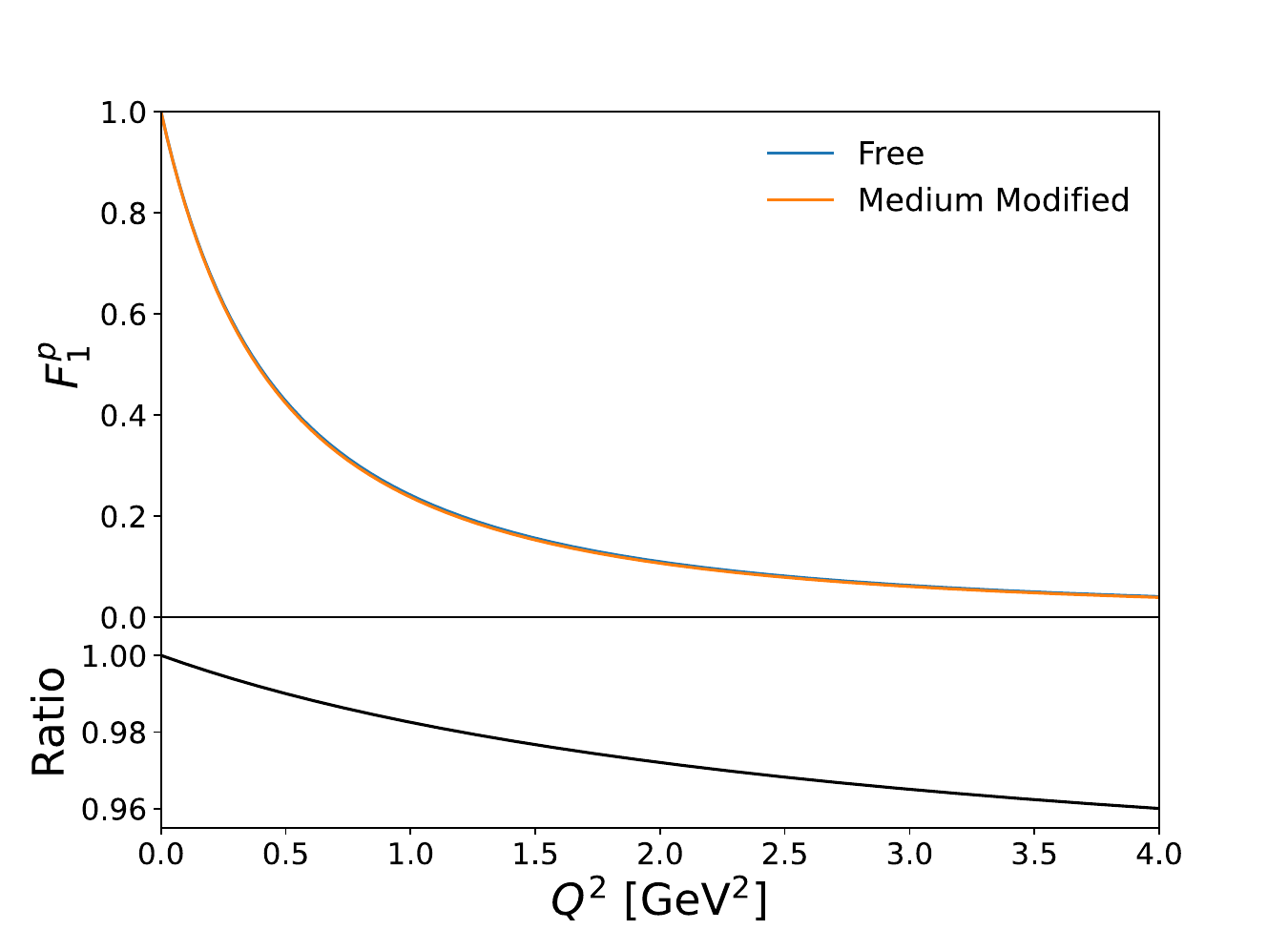}
  \hspace{0.3cm}
  \caption{(color online) (top)  Results for the $F_1$ elastic form factor of a free (blue) and medium modified (orange) proton in a Pb nucleus from Ref.~\cite{Kim:2022lng}. The bottom plot presents the ratio of the medium modified $F_1$ form factor to the free one.}
  %this is from finadeltashell.nb in NNuc
\label{Mmod}  \end{figure}

Why then, are   the computed   medium modifications of elastic electromagnetic form factors of Refs.~\cite{Kim:2022lng}    and ~\cite{Xing:2023uhj} so very different? This is because using nuclear DIS data does not unambiguously determine the light-front wave function of the nucleus. Even after starting with Ads/CFT duality there  still are many different ways to model the medium modification. The Drell-Yan \& West relation that relates lepton-nucleon  structure functions and form factors may be valid for nucleons, but its validity for nuclei is far less certain. A simple extension to nuclei would involve the largest possible values of $x$, that is $x\to A$.

The implication of the previous paragraph is that it would be very useful to obtain a direct experimental measurement of medium-modified form factors  It would  be especially useful for such measurements  to be made at kinematics where the nucleon virtuality is  large, because all current measurements correspond to small values of the virtuality. A discussion of different experimental approaches is contained in ~\cite{PhysRevC.99.035205}. See also~\cite{Melnitchouk:1996vp}.
         
The present note shows that it is possible to find a model that relies on short-ranged correlations which can account for the relevant facts, including also the plateaus observed in $(e,e')$ scattering from nuclei at values of $x>1$, an effect that is absent from models that rely solely on mean-field effects to reproduce the EMC effect. Ref.~\cite{Xing:2023uhj} frames the discussion of the EMC effect in terms of either mean-field or short-range correlations. 
Both mean-field  and SRC effects contribute to the average virtuality, roughly in the ratio $1:2$~\cite{CiofidegliAtti:2007ork}.
Indeed both effects must be involved as noticed long ago~\cite{Hen:2013oha}. The fundamental forces that bind nuclei involve two and three nucleons. The mean-field is obtained by averaging the fundamental forces over  densities. This is only an approximation. There must be residual effects of the basic forces, those that cause short-range-correlations. Much experimental data on nuclear DIS exists. It is now necessary  to obtain a consistent treatment of mean-field and short-range correlation effects.
        
   {\bf Acknowledgments}
The    work of DNK and GAM  was supported by the U.S. Department of Energy Office of Science, Office of Nuclear Physics under Award No. DE- FG02-97ER-41014.
 The research of MS was supported by BSF grant 2020115 and by US Department of Energy Office of Science, Office of Nuclear Physics under Award No. DE-FG02-93ER40771. The work of LW was supported by the U.S. Department of Energy Office of Science, Office of Nuclear Physics under Award No. 
 DE- FG02-96ER40960. The  work of EP was supported by  the  OPRA foundation, Pazy foundation, and the Israeli Science Foundation (Israel).  The work of OH was supported by  the U.S. Department of Energy Office of Science, Office of Nuclear Physics under Award No. DE-SC0020240.

\maketitle     
\noindent

\input{Replyv2.bbl}
% \bibliography{Reply2}{}
 
\end{document}

%% file: Replyv2.bbl
%merlin.mbs apsrev4-1.bst 2010-07-25 4.21a (PWD, AO, DPC) hacked
%Control: key (0)
%Control: author (0) dotless jnrlst
%Control: editor formatted (1) identically to author
%Control: production of article title (0) allowed
%Control: page (1) range
%Control: year (0) verbatim
%Control: production of eprint (0) enabled
%